\documentstyle[12pt,twoside,fleqn,espcrc1,epsfig]{article}

\newcommand{\AmS}{{\protect\the\textfont2
  A\kern-.1667em\lower.5ex\hbox{M}\kern-.125emS}}

\title{Mass shift, width broadening and spectral density of $\rho$-mesons
       produced in heavy ion collisions}

\author{V.L. Eletsky\address{Universit\"at Erlangen-N\"urnberg, 
        Erlangen, D-91058, Germany}%
        \thanks{On leave of absense from: ITEP, 117218 Moscow, Russia}
        ,
        B.L. Ioffe\address{ITEP, 117218 Moscow, Russia}
        and J.I. Kapusta\address{University of Minnesota, Minneapolis, 
        MN 55455, USA}}

\begin{document}
\maketitle

\begin{abstract}
Modifications of $\rho$-mesons formed at the last stage
of evolution of hadronic matter produced in heavy ion collisions are 
studied. It is found that while the mass shift is on the order of 
a few tens of MeV, the width and spectral density become so broad that
$\rho$ may lose its identity as a well defined resonance.
\end{abstract}

\section*{}
The problem of how the properties of hadrons change in hadronic or
nuclear matter in comparison to their free space values has attracted
a lot of attention. It is
clear on physical grounds that the in-medium mass shift and width
broadening of a particle are only due to its interaction with the
constituents of the medium, for not too dense media anyway.
Thus one can use phenomenological information
on this interaction to calculate the mass shift and width
broadening\cite{sh,ei}.

For meson $a$ scattering on hadron $b$ in the medium the contribution to
the self-energy is:
\begin{equation}
\Pi_{ab}(E,p) = - 4\pi \int \frac{d^3k}{(2\pi)^3} \,
n_b(\omega) \, \frac{\sqrt{s}}{\omega}
 \, f_{ab}^{(\rm cm)}(s) 
\end{equation}
where
$E$ and $p$ are the energy and momentum of the meson,
$\omega^2 = m_b^2 + k^2$,
$n_b$ is the occupation number,
and $f_{ab}$ is the forward scattering amplitude.
The normalization of the amplitude corresponds to the standard form of the
optical theorem 
$\sigma = (4\pi /q_{\rm cm}) {\rm Im} f^{(\rm cm)}(s)$.
The applicability of eq. (1) is limited to those cases where interference
between sequential scatterings is negligible.
In the limit that the target particles $b$ move nonrelativistically,
$\Pi_{ab} = -4\pi f_{ab}^{(b\,{\rm rest\,frame})} \rho_b$, 
where $\rho_b$ is the spatial density. This corresponds to the mass shift
and width broadening\cite{ei}
\begin{equation}
\Delta m_a (E) = 
-2\pi\frac{\rho}{m_a}{\rm Re} f_{ab}^{(b\,{\rm rest\,frame})}(E) \, , ~~~
\Delta \Gamma_a (E) = \frac{\rho}{m_a}~k\sigma_{ab}(E) \, .
\end{equation}

These relations hold also in the general case provided the amplitudes are 
averaged over momentum distributions of the constituents.  
We assumed\cite{us} that $\rho$-mesons are
formed during the last stage of the evolution of hadronic matter created in
a heavy ion collision, 
\newpage
when the matter can be considered as
a weakly interacting gas of pions and nucleons.
This stage is formed when the {\em local} temperature is on the order of
100 to 150 MeV and when the {\em local} baryon density is on the order
of the normal nucleon density in a nucleus.
The description of nuclear matter as such a gas, of course, cannot 
be considered as
a very good one, so it is clear that our results may be
only semiquantitative. The main ingredients of our calculation are 
on-shell $\rho\pi$
and $\rho N$ forward scattering amplitudes and total cross sections.
The scattering amplitudes were obtained by saturation of the 
low energy part with resonances and using a combination of
vector meson dominance (VMD) and Regge theory at high energy\cite{us}.
Using VMD we are restricted to the case of transversally polarized 
$\rho$-mesons. One can argue though, that for unpolarized $\rho$-mesons
our results should be multiplied by a factor ranging from
2/3 to 1\cite{us}.
Using experimental information on momentum distributions of pions
and nucleons and on the $\pi /N$ ratio shows that 
$\Delta m_{\rho} \sim$ tens of MeV, while the width increases by
several hundred MeV at beam energies of a few GeV$\cdot$A and by twice 
that amount at about a hundred GeV$\cdot$A.    

We also considered\cite{us1} the $\rho$ meson dispersion relation
for finite temperature and baryon density for momenta up to a
GeV/c or so as this is very interesting for the production of dileptons
in high energy heavy ion collisions. 
The dispersion relation is determined by the poles of the propagator
after summing over all target species and including the vacuum
contribution to the self-energy.
In the narrow width approximation we have
\begin{eqnarray}
E_R^2(p) &=& p^2+m_{\rho}^2+{\rm Re}\Pi_{\rho \pi}(p)
+ {\rm Re}\Pi_{\rho N}(p) \, , \nonumber \\
\gamma(p) &=& - \left[ {\rm Im}\Pi_{\rho}^{\rm vac} +
{\rm Im}\Pi_{\rho \pi}(p)
+ {\rm Im}\Pi_{\rho N}(p) \right]/E_R(p) \, .
\end{eqnarray}
where $E(p) = E_R(p) -i \gamma(p)/2$. We can also define a mass shift,
\begin{equation}
\Delta m_{\rho}(p) = \sqrt{m_{\rho}^2+{\rm Re}\Pi_{\rho \pi}(p)
+ {\rm Re}\Pi_{\rho N}(p)} - m_{\rho} 
\end{equation}
While $\Delta m_{\rho}$ in eq.(4) coincides with the one in eq.(1) for small
${\rm Re}\Pi/m_{\rho}$, the two definitions of width correspond to 
the rest frame of the $\rho$-meson (eq.(2)) and to that of the 
thermal system (eq.(3)) and differ by the time dilation factor, 
$\gamma=\Gamma m_{\rho}/E_R (p)$.

We evaluate $\Delta m$ and $\gamma$ for 
$T$ = 100 and 150 MeV and
nucleon densities of 0, 1 and 2 times normal nuclear matter density
(0.155 nucleons/fm$^3$).  This is done by utilizing
a Fermi-Dirac distribution for nucleons.  The nucleon chemical
potentials are 745 and 820 MeV for densities of 1 and 2 times
normal at $T$ = 100 MeV, and 540 and 645 MeV for densities of 1
and 2 times normal at $T$ = 150 MeV.  Anti-nucleons are not included.

Fig.1 shows the mass shift for different temperatures and nucleon 
densities (in units of normal nuclear density).
The effect with pions alone is negligible (on the order of 1 MeV).
The main effect comes from nucleons.
The effective mass increases with nucleon density and with momentum,
but is almost independent of temperature.
These trends and numbers are roughly consistent with other analyses
\cite{Wambach}.

Fig.2 shows the behavior of the $\rho$ meson width $\gamma (p)$.
Once again pions have very little
effect.  The main effect comes from nucleons.  Contrary to $\rho$
mesons moving in vacuum or through a pure pion gas the width
remains roughly constant with momentum when nucleons are present.
The width is about 240 MeV at 1 times nuclear density and about
370 MeV at 2 times nuclear density.  This means that the $\rho$ meson
becomes a rather poorly defined excitation with increasing
nucleon density.

\begin{figure}[htb]
\begin{minipage}[t]{75mm}
\epsfig{file=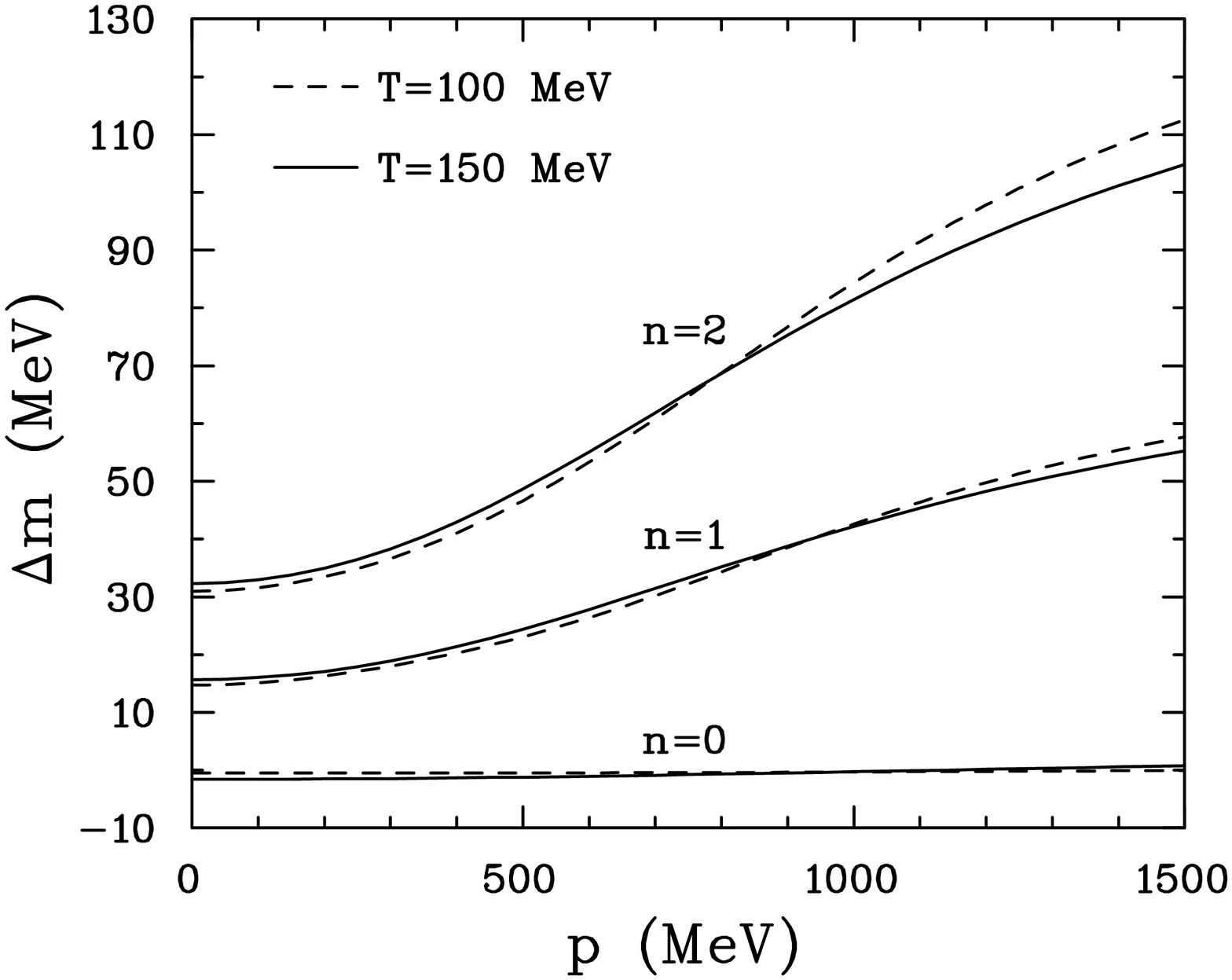,width=0.8\linewidth,angle=90}

\vspace{-1cm}

\caption{The $\rho$ meson mass shift (eq.(4)) as a function of momentum.}
\label{fig:largenenough}
\end{minipage}
\hspace{\fill}
\begin{minipage}[t]{75mm}
\epsfig{file=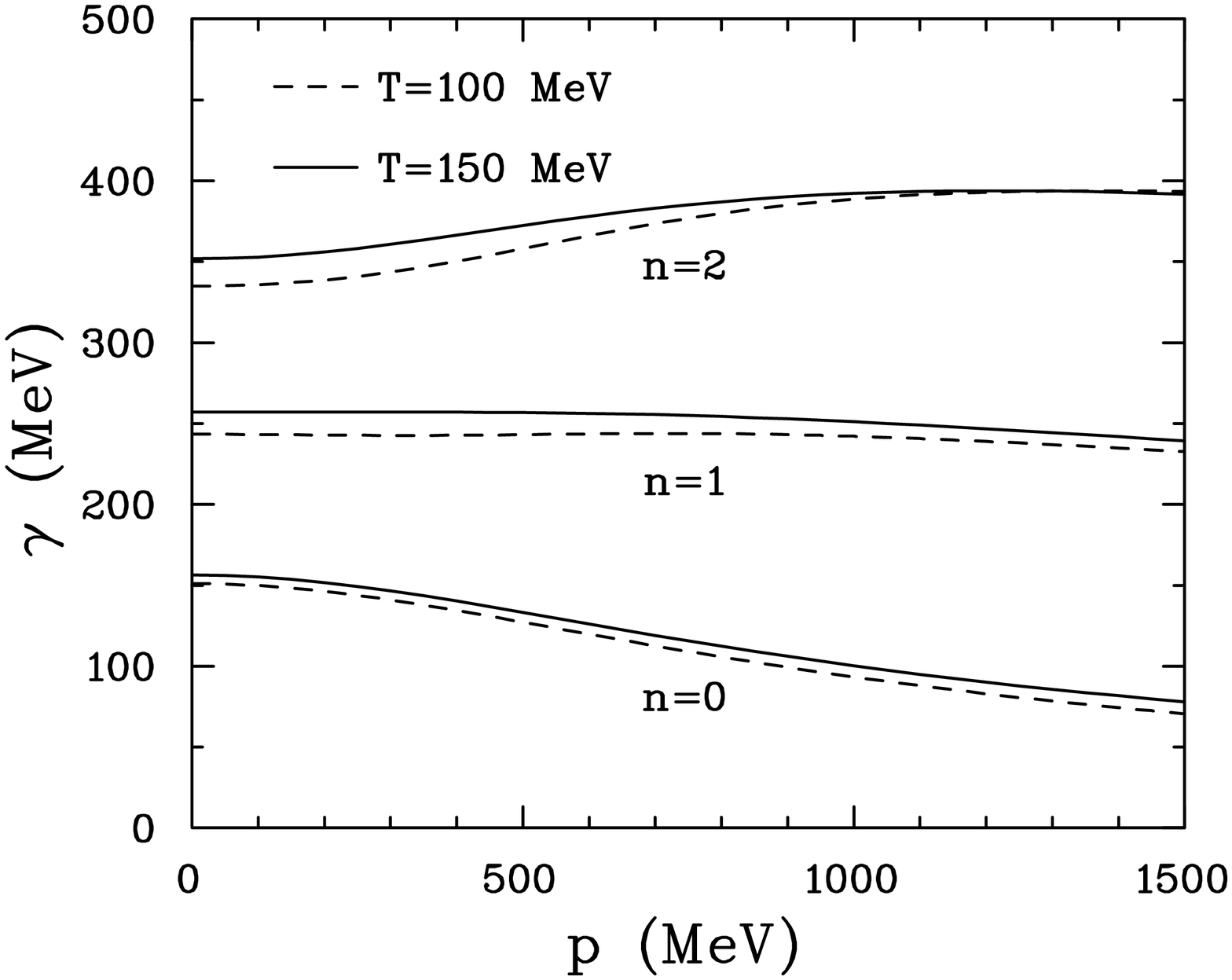 ,width=0.8\linewidth,angle=90}

\vspace{-1cm}

\caption{The $\rho$ meson width $\gamma(p)$ (eq.(3)) as a function of 
momentum.}
\label{fig:toosmall}
\end{minipage}
\end{figure} 

\vspace{-1cm}

\begin{figure}[htb]
\begin{minipage}[t]{75mm}
\epsfig{file=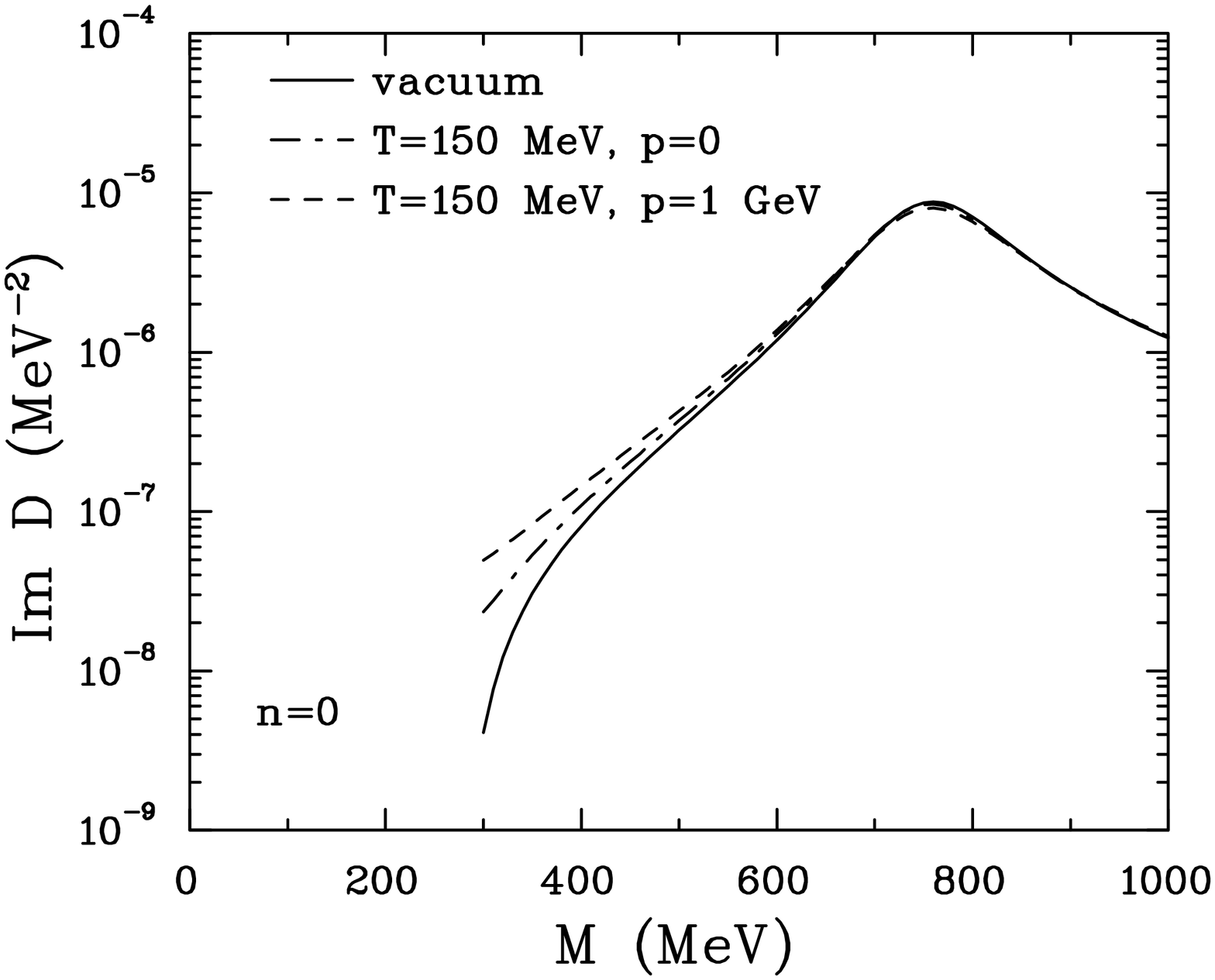,width=0.8\linewidth,angle=90}

\vspace{-1cm}

\caption{The imaginary part of the $\rho$ propagator. No nucleons.}
\label{fig:largenenough1}
\end{minipage}
\hspace{\fill}
\begin{minipage}[t]{75mm}
\epsfig{file=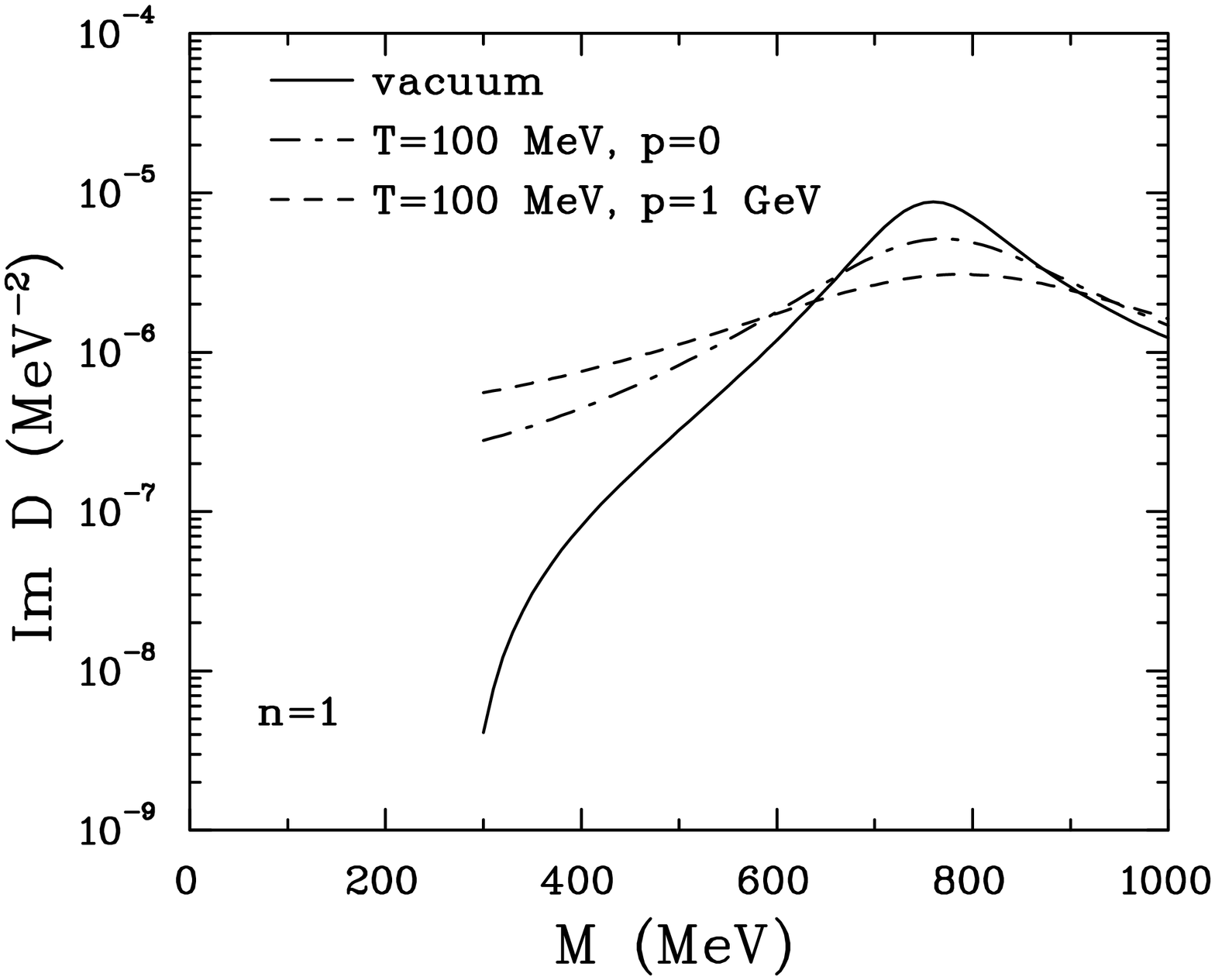,width=0.8\linewidth,angle=90}

\vspace{-1cm}

\caption{The imaginary part pf the $\rho$ propagator. With nucleons.}
\label{fig:toosmall1}
\end{minipage}
\end{figure}

\vspace{-0.5cm}

The rate of dilepton production is proportional to the
imaginary part of the photon self-energy \cite{mt} which is
itself proportional to the imaginary part of the $\rho$ meson
propagator because of VDM\cite{gk}.
\begin{equation}
E_+ E_- \frac{dR}{d^3p_+ d^3p_-} \propto
\frac{-{\rm Im} \Pi_{\rho}}{[M^2 - m_{\rho}^2 - {\rm Re} \Pi_{\rho}]^2
+ [{\rm Im}\Pi_{\rho}]^2}
\end{equation}
The vacuum part of $\Pi_{\rho}$ can only depend on the invariant mass,
$M^2 = E^2 - p^2$, whereas the matter parts can depend on
$E$ and $p$ separately.  Since we are
using the on-shell amplutudes, the matter parts
only depend on $p$ because $M$ is fixed at $m_{\rho}$. 
The vacuum parts are obtained from the Gounaris-Sakurai formula
\cite{gk}.  
The imaginary part of the propagator is proportional
to the spectral density.  The former is plotted in Fig.3 for
a pure pion gas and in Fig.4 for a gas of pions and nucleons at 
$T$ and $n$ characteristic of the final stages of a high energy heavy 
ion collision. Pions have very little
effect on the spectral density even at such a high temperature.
The effect of nucleons, however, is dramatic.  The spectral density
is greatly broadened, so much so that the very idea of a $\rho$
meson may lose its meaning.

The above remarks on the relative importance
of pions and nucleons may need to be re-examined when really
applying these calculations to heavy ion collisions, 
where $\pi /N \sim$ 6.  The pions seem to be overpopulated in 
phase space, compared to a thermal
Bose-Einstein distribution, and this could be modeled either
by introducing a chemical potential for pions or simply by
multiplication by an overall normalization factor.  Pions
would need to be enhanced by a substantial factor (5 or more)
to make a noticeable contribution at a density of 0.155 nucleons
per fm$^3$\cite{us}.

Recently data in Pb-Au collisions at 160 GeV$\cdot$A
have been presented \cite{data} where it was
found that the $\rho$-peak is absent at $k_T(e^+e^-)< 400$ MeV,
but reappears at $k_T(e^+e^-)> 400$ MeV.
This seems to be just the opposite of our findings.  However, our
calculations refer to the $\rho$ momentum relative to the {\em local}
rest frame of the matter and
a low momentum $\rho$ may actually be moving faster relative
to the outflowing matter than a higher momentum one.  No conclusion
can really be drawn without putting our results into a space-time
model of the evolution of matter.

In summary, we have studied the properties of the neutral
$\rho$ meson in the gas of pions and nucleons with experimental 
and thermal momentum distributions. In the former case, pions give the 
dominant effect, in the latter case they are not important.
This difference is due to completely different $\pi /N$ ratios.
However, in both cases interaction with the gas provides    
a generally positive mass shift for the $\rho$ mesons and
greatly increase their width.  The $\rho$ meson spectral density is so
broadened that the $\rho$ may lose its identity as a well defined
particle or resonance. At sufficiently high energy density
the matter can no longer be described very well as a gas
of noninteracting pions and nucleons.  Nevertheless the trends
must be obeyed by any realistic calculations of the $\rho$
meson in-medium.  Applications to thermal and hydrodynamic models
of heavy ion collisions are under investigation.

\section*{Acknowledgments}
The work reported here was supported in part by the RFBR grant
97-02-16131,  CRDF grant RP2-132 and the US 
Department of Energy grant DE-FG02-87ER40382.
V.L.E. thanks the Organizing Committee of QM99 for local support.



%
%




\end{document}